# On the Coupling between Ionic Conduction and Dipolar Relaxation in Deep Eutectic Solvents: Influence of Hydration and Glassy Dynamics

Aicha Jani,[†] Benjamin Malfait,[†] and Denis Morineau[†*]

[†]Institute of Physics of Rennes, CNRS-University of Rennes 1, UMR 6251, F-35042 Rennes, France

ABSTRACT

We have studied the ionic conductivity and the dipolar reorientational dynamics of aqueous solutions of a prototypical deep eutectic solvent (DES), ethaline, by using dielectric spectroscopy on a broad range of frequency (MHz-Hz) and for temperatures ranging from 128 to 283 K. The fraction of water in the DES was varied systematically to cover different regimes, starting from pure DES and its water-in-DES mixtures to the diluted electrolyte solutions. Depending on these parameters, different physical states were examined, including low viscosity liquid, supercooled viscous liquid, amorphous solid and freeze-concentrated solution. The ionic conductivity and the reorientational relaxation both exhibited characteristic features of glassy dynamics that could be quantified from the deviation from Arrhenius temperature dependence and non-exponential decay of the relaxation function. A transition occurred between the water-in-DES regime, (< 40 wt %), where the dipolar relaxation and ionic conductivity remained inversely proportional to each other, and the DES-in-water regime, (> 40 wt %), where a clear rotation-translation decoupling was observed. This suggests that for low water content, on the timescale covered by this study (~$10^{-6}$ s to 1 s), the rotational and transport properties of ethaline aqueous solutions obey classical hydrodynamic scaling despite these systems being presumably spatially microheterogeneous. A fractional scaling is observed in the DES-in-water regime, due to the formation of a maximally freeze-concentrated DES aqueous solution coexisting with frozen water domains at sub-ambient temperature.

**Corresponding Author:** * E-mail: **denis.morineau@univ-rennes1.fr**

1. Introduction

Recently, deep eutectic solvents (DESs) have received increased attention as an emerging class of solvents.[1-6] A majority of DESs are obtained by mixing of an H-bond donor (HBD) (typically an organic molecular component such as an amide, carboxylic acid, or polyol) with an H-bond acceptor (HBA) or with a molecular ionic species.[6] In most frequently studied systems, *i.e.* type III DESs, the latter consists of a quaternary ammonium salt. For compositions approaching the eutectic point a large decrease of the melting point is obtained, so that DESs form a liquid phase at room temperature.[7, 8] This phenomenon can be amplified by non-ideal mixing effects, which are favored by ionic and H-bonding interactions, and specific spatial correlations between the DES constituents.[9, 10] Nowadays, DESs are contemplated as cost-effective and sustainable alternatives to classical organic solvents.[3-6] They present many performant properties, such as low volatility, and high solvating and extracting properties. DESs have been considered for practical applications in a growing number of industrial processes, including extraction, (bio)synthesis, catalysis, electrochemistry, and carbon dioxide capture.[6]

The ionic conductivity of DESs makes them good candidates as alternative electrolytes for applications related to energy harvesting and storage.[6, 11-15] Recently, choline chloride-based DESs (i.e. ethaline, reline and aqueous solution of glyceline) have been studied as electrolytes for electrochemical applications,[11, 12] redox flow batteries,[13] lithium-ions batteries,[14] and dye-sensitized solar cells.[15]

From a fundamental point of view, it is striking that the ionic and molecular mobilities of DESs question classical theories. The rotational and translational dynamics of fluorescent dyes have been studied in reline,[16] ethaline,[17] and acetamide-urea DESs.[18] They exhibit fractional viscosity dependence and deviations from the hydrodynamic predictions based on Stokes-Einstein and Stokes-Einstein-Debye theories. These results suggest that the molecular motions of the fluorescent dyes are sensitive to the existence of dynamic heterogeneity in DESs. High frequency (GHz) dielectric relaxation (DR) have also concluded on the deviation from the hydrodynamic predictions for molecular rotation, indicating the critical role of collective reorientation relaxation and H-bond fluctuations in regulating the DR dynamics.[19, 20] From the combination of molecular dynamics (MD) simulation and neutron scattering (NS), a microscopic relation was made between transport properties and liquid structure of ethaline.[21] Different translational diffusion and rotational relaxation depending on the DES constituent were observed, and evidence for long-lasting dynamic heterogeneity (> 10 ns at 298 K) was inferred from the non-Gaussian character of the van Hove function related to self-diffusion of Chloride anion. Interestingly, dynamic heterogeneities were found to result from the complex interplay of solvation structure and H-bond dynamics.[21]

Aiming at a better understanding of the dynamics of DESs, it is worth noting that only a few studies have addressed longer time relaxation processes (i.e. longer than a few ns), so far.[22-24] Dielectric spectroscopy (DS) is a perfectly well-suited method because it allows independent measurements of the ionic conductivity and the orientation dipolar dynamics on an extended frequency range.[25] This provides a complementary view on the translation and rotational motions of DES species.[24] Moreover, it bridges the gap from short (~$10^{-9}$ to $10^{-6}$ s) to ultra-long timescale (~$10^2$ s), which are typical for the dynamical range going from a low-viscosity liquid to an amorphous solid. Due to the large depression of the eutectic melting point, DESs are usually liquid on a wide range of temperatures, and in frequent cases, eutectic crystallization can be avoided.

This further extends the slowing down of the DES dynamics in the supercooled and glassy states, implying that their dynamics should be studied over many decades.[24, 26] For ethaline, a comprehensive DS study has shown that the ionic transport and the reorientation dipolar motions were strongly influenced by the glassy dynamics.[24] In addition, it was demonstrated that both dynamics obeyed the same temperature dependence. The origin of this unique scaling has not been clarified yet. It might arise from a direct translational-rotational coupling, such as revolving-door mechanism or be due to the coupling of both dynamics to viscosity. The latter case would imply that hydrodynamics predictions from Stoke-Einstein and Stoke-Einstein-Debye models are fulfilled, thus minimizing the influence of dynamic heterogeneity.

In order to better understand the role of dynamic heterogeneity, an attractive approach would consist in varying gradually the solvation properties and H-bond interactions. Although conceivable in molecular models, this goal is more challenging in experiments. In the present study, we propose to use different hydration levels of DES as control parameters. In fact, water addition is recognized as a way to formulate DES-based solvents with tailored properties.[10, 26-37] More specifically, the structure of DESs is greatly influenced by water in several ways.[10, 27, 28, 36, 38, 39] Although supramolecular ionic clusters formed by the association of the initial DES components seem resistant to moderate hydration levels in the so-called 'water-in-DES' regime, water molecules often participate in the H-bond complexes of the original DESs as an additional HBD.[38, 40-42] The preferential solvation of chloride by water was also reported in choline chloride based DESs.[27, 33, 43, 44] Dielectric relaxation study of aqueous solution of glyceline and reline performed at room temperature and high (microwave) frequencies (0.05 to 89 GHz) indicates that a structural transition occurs for water mole fraction about 0.8, from the homogeneous electrolyte solution to a micro-heterogeneous structure with water-rich and DES-rich pools.[45] The micro-segregation of species into regions of different composition can result in co-continuous structures and mesoscopic length scale heterogeneities.[10, 35, 36, 39, 46]

In the present study, we have addressed the relative impact of these structural reorganizations on transport properties, dipolar relaxation and glassy behavior. We have performed a dielectric spectroscopy study of the prototypical ethaline DES (choline chloride/ethylene glycol 1:2) and nine different aqueous mixtures thereof, which were recently studied by differential scanning calorimetry (DSC) and neutron scattering.[26, 39] The ionic conductivity and reorientation dynamics were shown to vary by more than 6 decades on the entire studied temperature range 128 – 283 K. In the 'DES-in-water' region, (> 40 wt %), according to the designation introduced by Roldán-Ruiz *et al.*[36], an abrupt decoupling between transport properties and dipolar relaxation was observed. This was attributed to phase separation and the formation of a maximally freeze concentrated DES solution during cooling. In the 'water-in-DES' regime, (< 40 wt %), the solution obviously formed a macroscopically homogeneous liquid phase and presented characteristic features of a glassy dynamics, i.e. non-Arrhenius temperature dependence and non-exponential relaxation functions. In this range of composition, temperature, and timescale, this study indicates that only a marginal decoupling between ionic transport and dipolar dynamics is observed.

## 2. Methods

**2.1. Samples.**

Choline chloride (>99%) and ethylene glycol (anhydrous, 99.8%) were purchased from Sigma-Aldrich. DES was prepared by weighting and adding ethylene glycol and choline chloride in a molar ratio of 2:1, which is the accepted composition of the eutectic mixture.[47, 48] The DES were mixed by mechanical agitation at about 60°C for 30 min until a clear homogeneous liquid phase was obtained and served as stock solutions. Series of 10 working solutions were prepared by pipetting and addition of deionized water, corresponding to regularly spaced values of the mass fraction of water $W$ from 0 to 90 (%wt).

### 2.2. Dielectric Spectroscopy Experiments.

The sample was prepared in parallel plate geometry between two gold-plated electrodes with a diameter of 20 mm and a spacing of 260 µm maintained by Teflon spacers. It was placed in the cryostat and maintained under a dry nitrogen atmosphere. The complex impedance of the as-prepared capacitor was measured from 1 Hz to $10^6$ Hz with a Novocontrol high resolution dielectric Alpha analyzer with an active sample cell. The measurements were performed at thermal equilibrium with a temperature step of 5 K, and covering the temperature range from 283 K to 128 K (10 °C to -145 °C). The temperature of the sample was controlled by a Quatro temperature controller (Novocontrol) with nitrogen as a heating/cooling agent providing a temperature stability within 0.1 K.

## 3. Results and discussion

### 3.1. Dielectric permittivity spectra.

The complex dielectric function of the sample $\varepsilon^*(f) = \varepsilon'(f) - i\varepsilon''(f)$ was evaluated for the ten DES samples with different hydration levels, where $f$ denotes the frequency of the electric field, $\varepsilon'$ and $\varepsilon''$ the real and loss part of the complex dielectric function and $i$ symbolizes the imaginary unit. The two components are presented in Fig. 1(a) and (b) for a mixture of ethaline with $W = 10$ wt% of water. Qualitatively comparable spectra were obtained for all the samples. Depending on the temperature and the frequency ranges, different contributions were observed in the instrumental frequency window. At high temperature, the complex dielectric function was dominated by ionic conductivity. This was revealed by a power law term (-1/$f$) in the intensity of the dielectric loss (cf. blue dashed line with slope -1 in log-log scale in Fig. 1(b)). This ionic conductivity ultimately led to a huge increase of both $\varepsilon'$ and $\varepsilon''$ at lower frequency, with deviation of $\varepsilon''$ from the (-1/$f$) scaling. These phenomena were attributed to polarizations effects due to the blocking electrodes, as usually observed for conducting liquids. They do not hold any useful physical information on the samples and have been excluded from the discussion. In the low temperature range, typically below 210 K, a main peak appeared in the dielectric loss. It was also associated to a jump in the real permittivity $\varepsilon'$, as indicated by solid lines and red filled areas in Fig. 1(a) and (b). This was attributed to the dipolar α-relaxation process, in agreement with the interpretation made for anhydrous ethaline.[24]

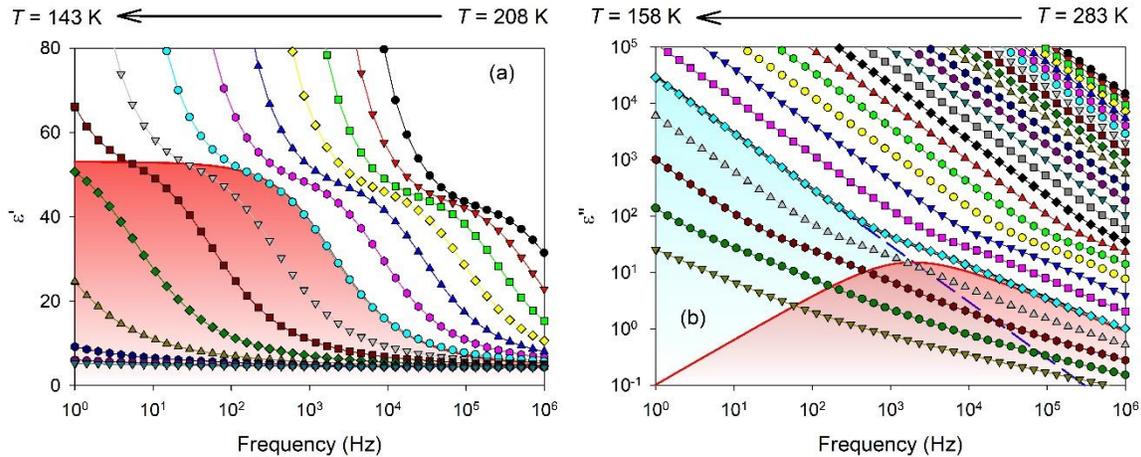

FIG. 1. (a) Real and (b) loss part of the complex dielectric function of a mixture of ethaline with $W = 10$ wt% of water as a function of the frequency and the temperature. The different curves correspond to temperatures varying (a) from 208 K to 143 K and (b) from 283 K to 158 K, with decreasing steps of 5 K from right to left. The components of the fitted model are illustrated for the selected temperature $T = 178$ K. The dipolar relaxation modelled by a Havriliak-Negami function (solid line with red-filled area in both panels) and ionic dc-conductivity (dashed line with blue-filled area in panel b).

The dielectric measurements were analyzed quantitatively by fitting the complex dielectric function by a model including a Havriliak and Negami functions (HN-model),[49] and a conductivity term according to eq. 1.

$$\varepsilon^*(\omega) = \varepsilon_\infty + \frac{\Delta\varepsilon}{(1+(i\omega\tau_{HN})^\alpha)^\beta} - i\frac{\sigma}{\omega\varepsilon_0} \qquad (1)$$

where $\varepsilon_\infty$ is the sample permittivity in the limit of high frequency, $\Delta\varepsilon$ and $\tau_{HN}$ are the dielectric strength and the HN-relaxation time of the mode. According to the formalism of the HN-model, the exponent parameters $\alpha$ and $\beta$ account for the symmetric and the asymmetric broadening of the complex dielectric function with respect to the Debye one. $\sigma$ stands for the DC conductivity of the sample and $\varepsilon_0$ the permittivity of vacuum. The application of this model provided fits of good quality in the temperature and frequency regions where electrode polarization was not significant, as illustrated in Fig. 1. The average relaxation time corresponding to the maximum frequency of the relaxation peak was computed according to

$$\tau = \tau_{HN} \sin\left(\frac{\pi a}{2+2\beta}\right)^{-1/\alpha} \sin\left(\frac{\pi a\beta}{2+2\beta}\right)^{1/\alpha} \qquad (2)$$

### 3.2. Ionic conductivity around room temperature

The ionic conductivities measured by fitting the complex dielectric function were found in good agreement with recent measurements reported in the complementary region above room temperature by Lapena *et al.*[34] These measurements were however limited to pure ethaline and one ethaline-water mixture (8 wt%) as illustrated in Fig. 2a. On increasing the amount of water at a constant temperature ($T = 283$ K), the conductivity gradually increased until it reached a maximum value for $W > 40\%$ as illustrated in Fig. 2b. This non-monotonous behavior has been also observed by molecular dynamics simulations of aqueous reline and ethaline solutions by Celebi *et al.*[50] The initial increase of conductivities with adding water was attributed to the destabilization of the H-bonds associations between DESs species that resulted in a smaller viscosity and faster transport properties. For $W > 40\%$, DESs constituents were found fully dissolved into a DES-in-water aqueous solution, which implied that the ionic conductivity eventually decreased due to dilution effects.

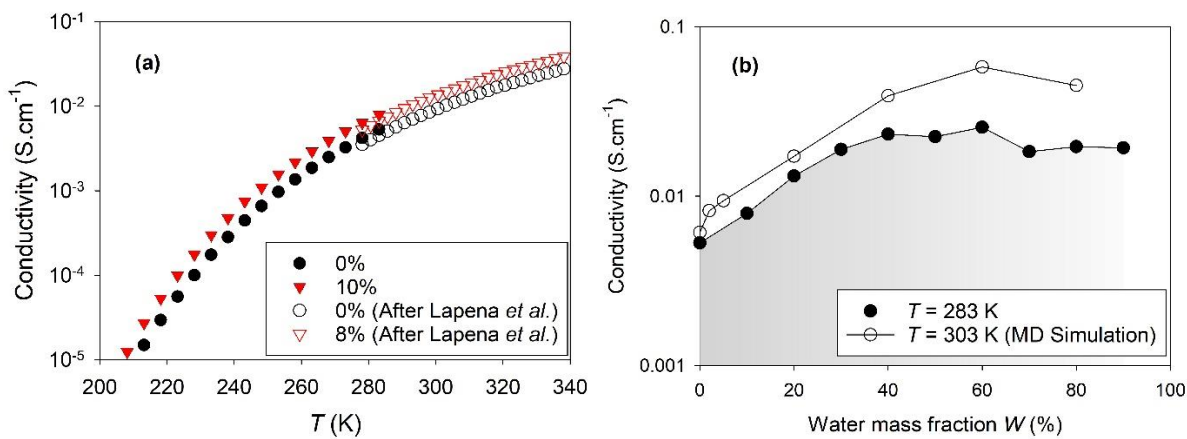

FIG. 2. (a) Comparison of the conductivity measured in the present study from the fit of the complex dielectric function of ethaline and an ethaline water mixture ($W = 10$ wt%) with the measurements performed above room temperature by Lapena *et al.*[34] for similar compositions. (b)

Effect of water on the conductivity measured in the present study at 283 K and by molecular dynamics simulation at 303 K by Celebi *et al*.[50]

### 3.3. Translational dynamics at variable temperature

The ionic conductivities measured on the temperature range from 158 to 283 K are illustrated in Fig. 3 in Arrhenius representation. Two qualitatively different behaviors were observed depending on the hydration level. For low hydration levels (0-30 %), a continuous temperature variation of the conductivity was obtained, while a discontinuity was observed for larger dilution (40-90%). The discontinuity occurred at a temperature that depended on *W*, as illustrated by an arrow in Fig. 3b for $W = 90\%$. These two ranges of compositions correspond to the usually denoted DES-in-water and water-in-DES regimes.[36]

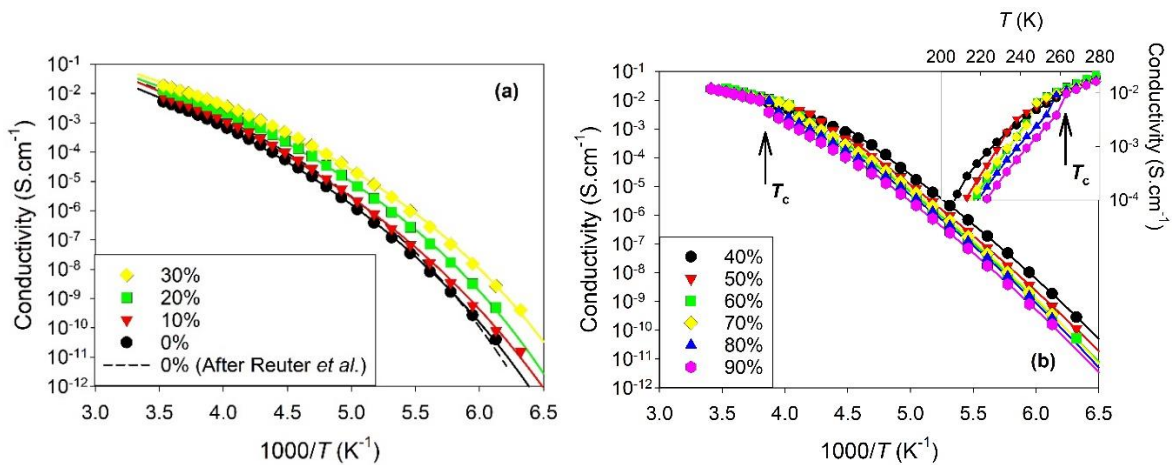

FIG. 3. Temperature dependence of the conductivity of aqueous mixtures of ethaline in Arrhenius representation for different values of the water mass fraction *W*. Fits with a VTF law (solid line). (a) Water-in-DES regime, with $W = 0\text{-}30\%$ (b) DES-in-water regime, with $W = 40\text{-}$

90%. The crystallization temperature $T_c$ is indicated by an arrow for $W = 90\%$. Inset: magnified view of the crystallization region with temperature in linear scale.

A clear explanation of this dependence on the water dilution can be found in the phase diagram of the ethaline-water system, which was recently determined by calorimetry (cf. Fig. 4).[26] It has been shown that below a threshold value $W_g' \sim 30\%$, water hydration molecules are strongly interacting with DES molecular units. In this range of composition, the hydrated DES forms a single phase system, even at low temperature, where it presents a single glass transition. On the contrary, for $W > W_g'$, the aqueous solution phase separates on cooling into a pure ice phase and a maximally freeze-concentrated aqueous solution of DES having the composition $W_g'$.

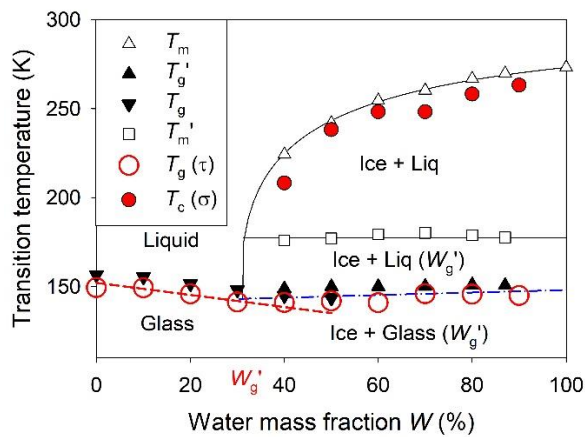

FIG. 4. Phase diagram of the DES aqueous mixtures as a function of the water mass fraction. Crystallization temperature determined from conductivity (filled circles), and isochronic glass transition temperature determined from dielectric relaxation time (open circles). Calorimetric determinations of the liquidus ($T_m$), onset of melting ($T_m'$) and glass transition temperatures ($T_g$ and $T_g'$) are extracted from Jani *et al.*[26]

For $W < W_g$', the conductivity monotonously increases with the fraction of water. This behavior follows the observations made previously around room temperature, and confirms the plasticizing effect of water. For $W > W_g$', the opposite dependence of the conductivity with the fraction of water is observed below $T_c$. This is due to the formation of a spatially heterogeneous phase separated system, where the ionic diffusion process that persists in the freeze-concentrated electrolyte is restricted by the coexisting ice matrix. Although the composition of the conducting freeze-concentrated electrolyte ($W_g$') does not depend on the initial composition $W$, the overall conductivity of the heterogeneous system decreases as a function of the relative fraction of ice formed.

### 3.4. Reorientational dynamics

The dielectric relaxation times are illustrated in Arrhenius representation in Fig. 5. As for the transport properties, two different behaviors were observed depending on the regime of hydration. For low hydration levels ($W < W_g$'), the dipolar relaxation became faster with the addition of water (cf. Fig. 5a). It was accelerated by up to a factor of 100 for the hydration level $W = 30\ \%$, with respect to the neat DES. The acceleration of the reorientational dynamics at low hydration level resembles the observation made for the conductivity. As general trends, these observations confirm that the addition of water to DES reduces viscosity and enhances molecular mobility.

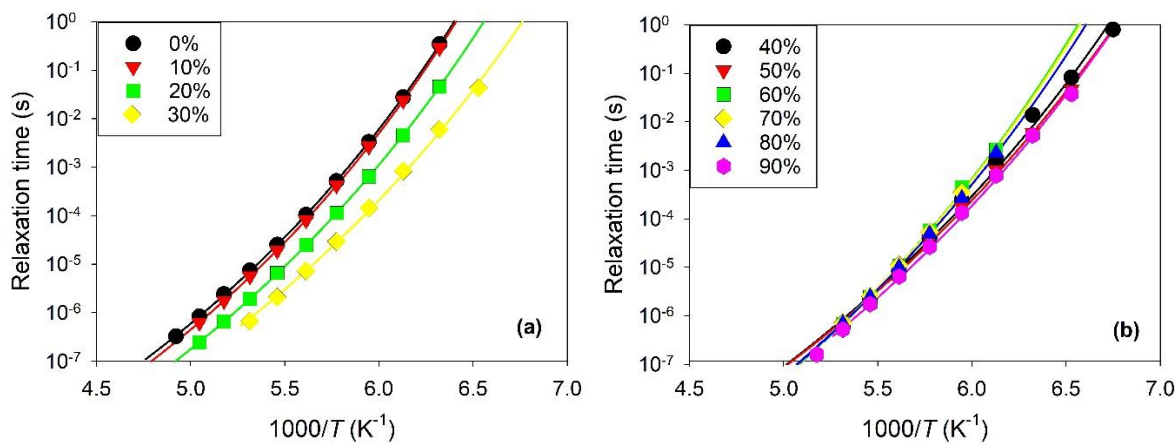

FIG. 5. Temperature dependence of the relaxation time of aqueous mixtures of ethaline in Arrhenius representation for different values of the water mass fraction $W$. Fits with a VTF law (solid line). (a) Water-in-DES regime, with $W = 0$-$30\%$ (b) DES-in-water regime, with $W = 40$-$90\%$.

For high hydration levels ($W > W_g$'), the dipolar dynamics depended only marginally on the amount of water. Indeed, the relaxation times measured for 6 different compositions essentially felt on a single curve, which extended over 7 decades in time (cf. Fig. 5b). This behavior differs from the ionic conductivity that decreases with $W$. This apparently contradictory observation finds a clear explanation. The dipolar relaxation probes the reorientational dynamics of the DES solution on a local scale. For $W > W_g$', the co-existence of ice with the phase forming the maximally freeze-concentrated solution of DES does not affect the rotational dynamics of the DES molecules. This implies that the size of the DES domains is large enough to limit the influence of interfacial effects and to prevent that the molecules can diffuse across a DES domain on the typical timescale of the rotational dynamics. On the contrary, these conditions may not be fulfilled for ionic conduction, which proceeds through translational diffusion of the ionic species.

### 3.5. Glassy dynamics

We now analyze in more details the functional temperature dependence of both dynamics. In all cases, a strong deviation from an Arrhenius behavior was observed. This is demonstrated by the curvature in the Arrhenius plots of the conductivity and the relaxation time shown in Fig. 3 and Fig. 5. This increase of the apparent activation energy during cooling is typical for glassforming liquids and it has been already discussed for neat ethaline.[24] For all the water-ethaline mixture studied, the existence of a glass transition has been actually by demonstrated by DSC.[26] A complementary estimation of the glass transition temperature has been also evaluated from the present dielectric measurements by extrapolation of the relaxation time to $10^2$ s as illustrated in Fig. 5.

The deviations of the dipolar relaxation and conductivity from the Arrhenius behavior were described by the empirical Vogel–Fulcher–Tammann (VFT) law. For the relaxation time, the classical VFT equation is given by Eq. 3, while the modified version corresponding to Eq. 4 is classically adopted for the ionic transport.[24]

$$\tau = \tau_0 exp\left(\frac{D_\tau T_{0\tau}}{T-T_{0\tau}}\right) \quad (3)$$

$$\sigma = \sigma_0 exp\left(\frac{-D_\sigma T_{0\sigma}}{T-T_{0\sigma}}\right) \quad (4)$$

In these two equations, $\tau_0$ and $\sigma_0$ are the pre-exponential factors, $D_x$ and $T_{0x}$ ($x = \tau, \sigma$) are the strength parameters and the Vogel-Fulcher temperatures associated to the dipolar relaxation and conductivity, respectively. Large deviation from an Arrhenius behavior is characterized by small values of $D_x$, while an Arrhenius behavior is recovered for large $D_x$ and $T_{0x} \approx 0$. Satisfactory fits of the experimental relaxation times and conductivities were obtained for the studied range of temperature and water mass fraction as illustrated by solid lines in Fig. 3 and Fig. 5. The deviation of the relaxation time from an Arrhenius behavior has been considered for many decades as a salient feature of most of the glassforming systems in their liquid and supercooled states. It has been recognized as an evidence for the cooperative nature of the structural relaxation.[51] The non-Arrhenian character has been used to define a parameter, named the fragility index $m$, that categorizes series of glass-forming liquids according to the nature of their dynamics in terms of fragile or strong systems.[52] For ethaline, we found values of the fragility index around $m = 48$, which corresponds to the type of intermediate system according to this classification. For pure ethaline, Reuter *et al.* found a slightly larger value of the fragility index ($m = 60$), which might be attributed to the larger frequency range adopted in that study ($10^9$-$10^{-1}$ Hz). Interestingly, we observed that the fragility of ethaline decreased with the addition of water, as illustrated in Fig. 6, and then reached a plateau value ($m = 40$) for water mass fraction larger than $W = 40\%$. The invariance of $m$ in the DES-in-water regime, i.e. for $W > 40\%$, agrees well with the formation on cooling of a maximally freeze-concentrated solution with fixed composition $W_g$' and dipolar dynamics. In the water-in-DES regime, the possible reduction of fragility of a DES with adding water has never been reported so far, though it can be seen as moderate, if one considers that the entire fragility scale extends from $m = 16$ for strong liquids to $m = 170$ for extremely fragile systems.

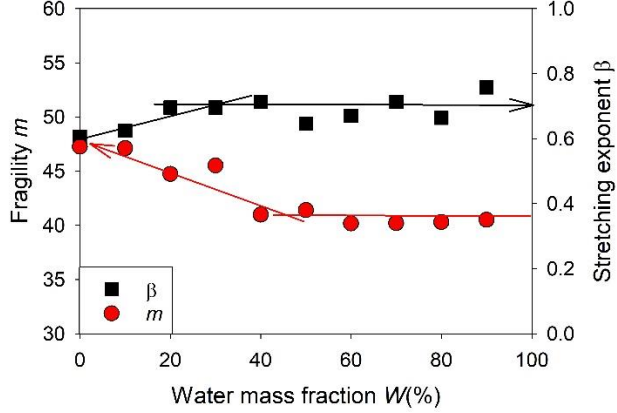

FIG. 6. Dependence on the water amount of the fragility index (circle) and the stretching exponent of the dipolar relaxation (square).

Another salient observation concerns the deviation of the relaxation function from a simple Debye process. In the time domain, the non-exponential character of the relaxation function of liquids is often expressed by a stretched exponential function, also denoted Kohlrausch-William-Watts function. To conform with this representation, we applied the numerical ansatz that connects the stretching exponent $\beta_{KWW}$ with the HN parameters obtained in the frequency domain i.e. $\beta_{KWW} = (\alpha\beta)^{\frac{1}{1.23}}$.[53] The found values of $\beta_{KWW}$ increased from 0.6 to 0.7 with the addition of water, and saturated for mass fraction larger than $W = 40\%$, as illustrated in Fig. 6. In glass forming liquids, the non-exponential character of the structural relaxation is classically attributed to the presence of a distribution of relaxation times. There is a growing number of evidences that this distribution actually reflects the presence of spatially heterogeneous dynamics.[53] Accordingly, it seems counterintuitive, that the addtion of water to DES resulted in apparently more homogeneous reorientational dynamics. From a structural point of view, the addition of water as a competing

HBD and solvating agent with respect to the initial DES constituents rather point to the formation of a broader distribution of local environments.[10, 27, 35, 39, 46] Therefore, one would have expect a decrease of $\beta_{KWW}$ with increasing $W$. A possible rationalization is suggested by the concomitant reduction of fragility of the DES with the addition of water. As shown in Fig. 6, the increase of $\beta_{KWW}$ seems correlated to the decrease in the fragility index $m$. A relationship between fragility and non-exponentiality has been established for numerous very different glass-forming systems.[52] It appears that ethaline and water-ethaline mixtures also obey with this relationship between $\beta_{KWW}$ and $m$. Their dynamics compare well with other H-bonded liquids, such as polyalcohols, that fall in the category of intermediate and moderately non-Debye liquids (cf. Fig. 3 in ref. 52). Moreover, when adding water the increase of $\beta_{KWW}$ with decreasing $m$ is in qualitative agreement with the linearly decreasing dependence, which was proposed for molecular glassfomers in that study.[52] Supporting pioneering ideas about the concept of cooperativity dynamics,[51] many studies indicate that the glassy slowdown is accompanied by the growing length scale of transient domains.[54, 55] In this context, one may speculate that the nanosegregation of DES components with the addition of water, while favoring structural heterogeneity with the formation of domains having distinct composition, would restrict the growth of dynamic cooperativity to values bounded by the domains typical size. [35, 39] A microscopic description of cooperative character of the molecular dynamics in DESs would be desirable to test this idea, which shares similarity with the concept cut-off on cooperativity length introduced for nanoconfined liquids.[56-58]

### 3.6. Rotation-Translation decoupling

In order to address the (de)coupling of the ionic transport and the reorientation dipolar motions we have compared in Fig. 7 the strength parameter evaluated from VTF fits of both dynamical processes. Same values are obtained for neat ethaline, which is in agreement with the recent study concluded that both dynamics obeyed the same temperature dependence.[24] A difference between both strength parameters is observed on adding water. It remains very small up to $W = 40$, indicating opposite effect of water on the fragility when refereeing to reorientation or translation.

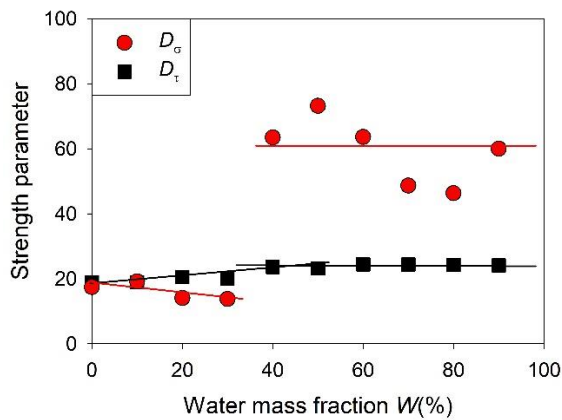

FIG. 7. Dependence on the water mass fraction of the strength parameter of the dipolar relaxation (circle) and the ionic conductivity (square) of ethaline aqueous mixtures.

It should pointed out that the frequency and temperature range covered by both processes are different in DS experiments, which might influence the determination of the strength parameters. To overcome the question, we have also represented the conductivity as a function of the relaxation time for a limited selection of temperatures where both processes are measurable simultaneously in the DS spectra, as illustrated in Fig. 8a. It confirms that for $W < 40$ %, and in this temperature range, both quantities are essentially inversely proportional. This relationship is indicated by a solid line in Fig. 8a. For neat ethaline, this inverse proportionality has been already demonstrated.[24] In this study, the authors concluded that the question remained opened, however, whether the two processes were actually coupled to viscosity, implying that the Stokes-Einstein and Stokes-Einstein-Debye relationships were fulfilled, or that both processes were directly coupled to each other. The latter case was illustrated by the revolving-door or paddle-wheel mechanism, which was also considered for ionic liquids where paths for ions diffusion are opened by the rotation of other dipolar ions.[59] The addition of water to choline based DES is known to modify significantly the hydration shell of chloride anions, which are preferentially solvated by water.[27, 33, 43, 44] The H-bond interactions within the original DESs supramolecular complexes are also affected by water acting as an additional HBD.[38, 40-42] For these reasons, it seems unlikely that a persistent revolving-door mechanism can tolerate the effects of adding up to 40% of water to ethaline. It rather indicates a more classical scenario in which both dynamics are coupled to viscosity, in accordance with the predictions from hydrodynamics.

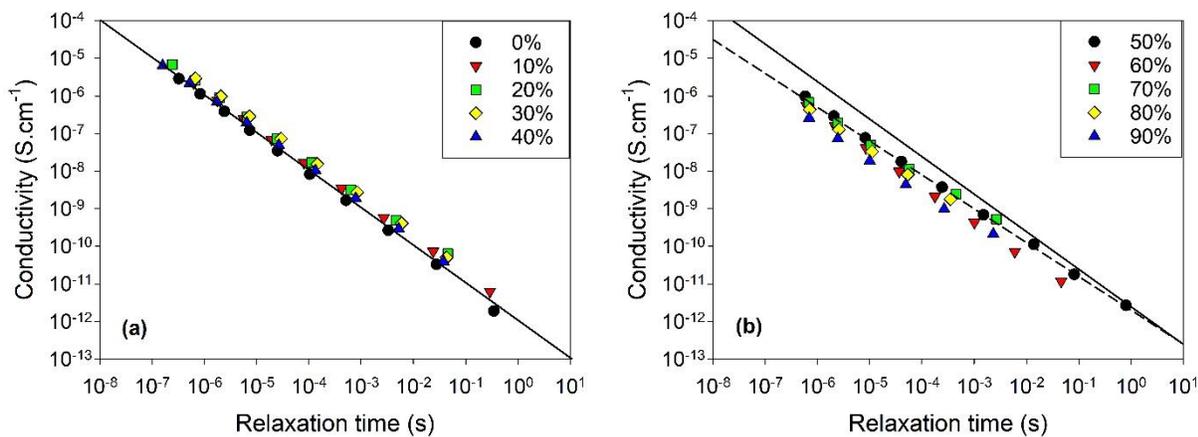

FIG. 8. Dependence of the ionic conductivities on their reorientational relaxation times of aqueous mixtures of ethaline as a function of the water mass fraction. (a) Water-in-DES regime, with $W$ = 0-30% (b) DES-in-water regime, with $W$ = 40-90%. The solid line with slope -1 in panels (a) and (b) indicates inverse proportionality $\sigma \propto \tau^{-1}$, while the dashed line in panel (b) has a slope -0.9, indicating $\sigma \propto \tau^{-0.9}$

For large water mass fraction ($W > 40\%$), a completely different situation is observed, with a large decoupling between both dynamics. This is illustrated by a sudden jump of the strength parameter of the ionic conductivity from about $D_\sigma = 15$ to 60 at $W = 40\%$, which is consistent by the close to Arrhenian behavior of the conductivity shown in Fig. 4b. Likewise, a breakdown of the inverse proportionality is observed, indicating a fractional translational-rotational coupling with $\sigma \propto \tau^{-0.9}$, as illustrated in Fig. 8b. This phenomenon provides a clear illustration of the possible effects of strong spatial heterogeneities, the latter being due to the coexistence of domains formed by maximally freeze-concentrated DES aqueous solutions and by solid water.

## 4. Conclusion

The existence of dynamic heterogeneity in DESs has been indicated by clear deviations of the rotational-translational dynamics of dye molecules from the Stokes-Einstein-Debye hydrodynamic predictions. The dielectric spectroscopy is a unique method to determine simultaneously the reorientational dynamics of dipolar molecules and the transport properties of ionic species on an extended range of frequency. So far, this method has been barely used for DESs, although a recent study has pointed to the existence of a coupling between both processes for ethaline.

In the present study, we have used the fraction of water in ethaline DES aqueous mixtures, as a varying control parameter of the solvation and H-bond interactions. The reorientational dynamics and the ionic conductivity of ten ethaline solutions were studied in the temperature range 128 – 283 K and the MHz-Hz frequency range. Two different regimes were observed depending on the water mass fraction $W$, which agrees with the published phase diagram of ethaline aqueous solutions.

In the water-in-DES regime, (< 40 wt %), the DES aqueous mixtures did not formed classically phase separated systems. The relaxation time and the conductivity presented characteristic features of glassforming systems, including a deviation from Arrhenius temperature dependence and a non-exponential decay of the relaxation function. The addition of water resulted in an acceleration of both dynamics, by up to a factor of 100, indicating a plasticizing effect of water on the DES. Moreover, according to the dipolar relaxation, the addition of water concomitantly reduced the fragility index $m$ and increased the stretching exponent $\beta_{KWW}$. According to the values of both parameters, ethaline and ethaline-water mixtures fall in the category of intermediate glassforming liquids. The values of $\beta_{KWW}$ and $m$ fulfills the relationship made for intermediate molecular glassforming liquids such as polyalcohols.

Interestingly, it can be concluded that in this range of composition, temperature, and timescale, the dipolar relaxation and ionic conductivity remained inversely proportional to each other. The persistence of a direct coupling mechanism, where paths for ions diffusion are opened by the rotation of other dipolar ions is improbable. Although such paddle-wheel mechanism was indicted for ionic liquids, and also invoked as a possible process for neat ethaline, it seems unlikely that it remains unaffected despite the important modifications in ion solvation and H-bond interaction induced by the water addition to the DES components. The applicability of the Stokes-Einstein-Debye relation rather indicates that both dynamics are equally coupled to viscosity. This conclusion also implies that the dynamic heterogeneity plays a secondary role in the present experiment situation.

In the DES-in-Water regime, (> 40 wt %), a clear rotation-translation decoupling is observed, including a transition towards a nearly Arrhenius dependence of the ionic transport, and a fractional relation between conductivity and dipolar relaxation with $\sigma \propto \tau^{-0.9}$. This behavior is attributed to the spatially heterogeneous nature of the phase-separated sample, where domains formed by maximally freeze-concentrated DES aqueous solutions coexist with frozen water.

## Acknowledgments

Support from Rennes Metropole and Europe (FEDER Fund – CPER PRINT$_2$TAN), and the ANR (Project NanoLiquids N° ANR-18-CE92-0011-01) is expressly acknowledged. This work is part of the PhD thesis of A.J. who benefits from a grant from the French Ministry of Higher Education, Research, and Innovation. The authors are grateful to the CNRS – network SolVATE (GDR 2035) for financial support and fruitful discussions.

## Data Availability

Derived data supporting the findings of this study are available from the corresponding author upon reasonable request.